\documentclass[11pt]{article}

\usepackage[T1]{fontenc}
\usepackage{lmodern}        
\usepackage[utf8]{inputenc} 

\usepackage[english]{babel}

\usepackage[letterpaper,top=2cm,bottom=2cm,left=3cm,right=3cm,marginparwidth=1.75cm]{geometry}

\usepackage{amsmath,amssymb,amsthm}
\usepackage{graphicx}
\usepackage{subcaption}

\usepackage[colorlinks=true,
            linkcolor=blue,
            citecolor=blue,
            urlcolor=blue]{hyperref}

\usepackage[numbers,sort&compress]{natbib}

\usepackage{booktabs}
\usepackage{array}
\usepackage{caption}
\usepackage{subcaption}
\usepackage{url}

\usepackage{algorithm}
\usepackage{algpseudocode}

\title{Hierarchy Entropy Degeneration Explains the Rat Utopia Population Collapse: The Role of Full Visibility and Isolation}
\author{Alexander V. Mantzaris}

\date{%
  Department of Statistics and Data Science, University of Central Florida, USA\\[1ex]%
  \today
}

\begin{document}
\maketitle

\begin{abstract}
Calhoun's Rat Utopia experiments demonstrated a puzzling population trajectory: initial growth, plateau, and eventually a total collapse of the rat population despite abundant resources. This paper proposes a hypothesis that the enclosure's design enabled full visibility of the social hierarchy (pecking order), leading to entropy degeneration: progressive loss of uncertainty in rats' perceived ranks over generations. High initial uncertainty drives engagement in dominance, reproduction, and care; as visibility solidifies the hierarchy over the generations, uncertainty vanishes, nullifying perceived gains from social activities. Simulations reproduce the experimental arc which rely on a game theoretic matrix that is parameterized by the uncertainty (entropy) in the hierarchy which changes over rat generations.
\end{abstract}

\section{Introduction}

The famous Calhoun Rat Utopia experiments \cite{calhoun1962population,calhoun1973death} studied a population of rats that were placed into an environment where all their needs for survival were provided for. The initial population of rats could socialize, feed and reproduce freely. In terms of survival ingredients it satisfies all the requirements. Studying the populations though, produced an unexpected trajectory for the population, there was an initial population growth that followed an expected organic linear increase on a semi-log axis but then afterwards it was followed by a plateau (decreased growth rate) and then a period of decline. The decline continued until the total population perished. The population had never reached the full capacity of the area either (at all times below 3840) or even came close to full capacity. Various theories exist as to why this occurred and some of these will be mentioned, but this work puts forward a different hypothesis to explain the trajectory.

This work puts forward the idea that the Calhoun experimental environment for the rats provided a clear view of the dominance hierarchy (pecking order) across all the rats so that after this hierarchy establishment was known the lack of uncertainty removed any attempts to engage in activities that touch upon that hierarchy. Effectively that the dynamics of all the rat engagements with their peers, environment or even their own well-being is parameterized by the entropy of their perceived position in the hierarchy. That a view of a static hierarchy nullifies the perceived gains of any kind of engagements. Using this paradigm a payoff matrix will be formulated and applied in a game theoretic simulation which shows a similar trajectory to the original experimental results. 

The original results (also known as Universe~25) are shown in Fig.~\ref{fig:universe25_original}, exhibits the canonical three-phase pattern: an early, approximately exponential expansion (Phase~B, Growth), a peak and short plateau where reproduction effectively stops despite high headcount (Phase~C, Plateau), and a prolonged decline toward extinction in the absence of new litters (Phase~D, Decline). Quantitatively, the population rises by DAC~315, peaks near \(N\!\approx\!2200\) around DAC~560 (coincident with the last recorded births), and then falls from \(N\!\approx\!2056\) at DAC~736 to very low levels over the ensuing \(\sim\!776\) days. The figure also notes an "optimal" adult level of roughly \(150\), emphasizing that the system overshot any steady capability before collapsing, with the log-scaled trajectory highlighting near-exponential growth and, later, near-exponential decay.
\begin{figure}[htbp]
  \centering
  \includegraphics[width=\linewidth]{plots/Mice_population_Universe_25.jpeg}
  \caption{\textbf{Universe 25: population trajectory and phase landmarks (Calhoun).}
  The ordinate shows population size \(N\) (log scale) and the abscissa is days after colonization (DAC) with 4 breeding pairs. 
  \emph{Phase B (Growth)}: a near-exponential rise from the founder group to \(N\approx 620\) by DAC~315. 
  \emph{Phase C (Plateau/Peak)}: continued increase to a peak \(N\approx 2200\) by DAC~560; births cease around the peak (\emph{“Last Born, 1~Mar~1970”}), after which the colony persists without replacement. 
  \emph{Phase D (decline)}: population declines from \(N\approx 2056\) at DAC~736 along an extended decay, approaching near-extinction by \emph{1~Mar~1972}; the annotation “776 days” highlights the interval from last birth to the late-stage census. 
  The marked band “\(N=150\) Optimum No. of Adults” indicates a putative sustainable adult count, underscoring that the observed peak represents a substantial overshoot preceding collapse. (image used from Wikimedia: \url{https://commons.wikimedia.org/wiki/File:Mice_population_Universe_25.jpg}}
  \label{fig:universe25_original}
\end{figure}

Before showcasing this new model, it is recommended that the reader at least look over the section~\ref{app:PD} in the Appendix. It is dedicated to showing how symmetric visibility in the Prisoner's dilemma game changes the expected payoffs in both the one-shot and the iterative games. Even in the prisoner's dilemma game visibility decreases the expected reward. This is used as inspiration for creating a parameterized game matrix that uses the uncertainty in the hierarchy over time to change the payoffs.

A set of key modeling assumptions are made to simplify the model and explore the core dynamics. It is assumed that there are 3 tiers in the hierarchy of the experiment, the dominant rats in the central area, the intermediates that reside on the periphery of the center along the corridors and the subordinates (aka 'the beautiful ones') that reside on the walls (rats in all areas had immediate access to resources close them in these areas). Another assumption is that the original rats inserted into the environment had no information about their position in the hierarchy. Their prior probability for their expected position is $p=1/3$ (uniform across each of the 3 tiers), and the first rat offspring would still have a $p=1/3$, but the second generation of offspring a probability $p=1/2$ and the third generation of offspring (last offspring before collapse) a probability of $p=1$; no uncertainty and zero entropy of their position in the hierarchy. The Appendix in section~\ref{app:EquationModel} showcases a differential equations based model that can roughly demonstrate the overall structure and pattern of the population trajectory of the experiment using these core assumptions. Although the focus is on the game matrix approach the methodology for using an agent based modeling simulation is also presented in the Appendix in section~\ref{sec:ABM}.
The code used in this study is written in the programming language Julia Lang and is hosted on Github at, \url{https://github.com/mantzaris/RatUtopia}.

\subsection{Background Literature}

A great deal of insight can be attained from a later paper of Calhoun \cite{calhoun1977looking} that provides insight in the experiment. The article of \cite{ramsden2009escaping} provides some interesting details into the history of the experiment but also discusses the caution that should be exercised in generalizing the results into the human context.
A key question that has arisen is that the collapse and behaviors that emerged with that phase stem from overcrowding. The work of \cite{marsden1972crowding}, discusses crowding but an argument against this exists since it is expected that for over crowding to produce complete collapse it would be expected that rats had exhausted the environment capacity but that is not the case.

A concise review that clarifies the shift from 'density per se' to social interaction and control of space; notes inconsistent human crowding emphasizing design/role structure over headcount \cite{ramsden2009urban}. The structure and design concept underlies the assumption made here as well since for the hierarchy to be established and symmetrically known to all rats the visibility requires structural support of the enclosure. The results of the experiment stand out since the structure of the enclosure also stands out as a unique area to facilitate this hierarchy information to all residing rats according to these assumptions.

Another common explanation is that there is general stress in these rats \cite{beery2015stress} placed in such a situation. There are arguments against this though since rats in the wild can have stress induced from predators and lack of resources. In the wild, areas of rat dwellings also show that rats exhibit acts of violence to establish dominance. Collections of rats in other experiments do not display the type of total collapse and emergence of behaviors like that of the beautiful ones based on the literature covered. 

There are various modeling efforts presented to simulate rat behaviors such as this generic stochastic model, \cite{perony2012stochastic}. Modeling rat groups \cite{chen2025modeling} uses a different paradigm. An interesting part related to the female rat psychology when exposed to uncertainty in male relationships (such as high turnover rates), Bruce effect, \cite{eccard2017bruce}. A relatively recent paper presents a dynamic systems modeling approach with elements of social physics by drawing analogies to the Ising model; \cite{chen2023generalized}.

In relation to the concepts of hierarchies in animals a key concept is the linearity in animal hierarchies \cite{appleby1983probability}. It is also found that in the networks of dominance networks motifs can be found, \cite{shizuka2015network}. Social rank can even be found in crayfish, \cite{hock2006modeling}. 
Very importantly the work of \cite{elwood2009difficulties} demonstrates how there is contest logic and assessment models which underlie how fights reveal resource‑holding potential (RHP). This can provide insight into rank dynamics and the perception of resources that are obtained from those outcomes. 
The work of \cite{dall2005information} shows a broad view of when animals sample, ignore, or socially acquire information, and how uncertainty and sampling costs shape adaptive behavior as it is assumed in this work. 
\cite{bromberg2009midbrain} shows how macaques pay to reduce uncertainty: they can sacrifice rewards to obtain advance information or for counterfactual outcomes, and willingness‑to‑pay scales with Shannon entropy. Midbrain dopamine neurons signal preference for information—highlighting that uncertainty reduction itself can invigorate behavior. 
In \cite{kraeuter2018open} a novel (hence uncertain) arena, rodents typically show transient hyper‑locomotion and rearing; a classic 'novelty‑induced' exploratory response—before settling. This is foundational in anxiety/locomotion assays and illustrates vigorous sampling under uncertainty. There is literature that supports that attention cues can alter reward chemicals in neuron pathways, \cite{angela2005uncertainty}.

Over generations the hierarchy can be reinforced by a process of transgenerational plasticity, \cite{donelson2018transgenerational}.
Another interesting approach is based on the self‑organizing hierarchies with reinforcement is where interaction outcomes update propensities, yielding linear/transitive hierarchies without central control how winner–loser plus spatial structure suffices to reproduce realistic steepness and linearity \cite{clutton1986great}. The reinforcement can also be viewed from a game approach, \cite{broom2002modelling} gives a unified model connecting formation and maintenance phases (hierarchy formation as a multi‑player Hawk–Dove game). 

The work of \cite{van2022hierarchical} demonstrates DomWorld, where dominance hierarchy, behavioral dynamics and network triad motifs in the model using analytical methods from a previous study on dominance in real hens is shown. It allows an exploration for the dynamics which underline how a linear dominance hierarchy emerges.

\section{A hierarchy–uncertainty parameterized matrix model of Calhoun's experiment}
\label{sec:methods}

\subsection{Time scale and phases}
Time advances in discrete monthly steps \(t=0,1,\dots\). Aligning key calendar landmarks with the Calhoun timeline as follows (months since enclosure start): onset of \emph{plateau} at \(m_{\mathrm{plateau}}=10.5\), onset of \emph{decline} at \(m_{\mathrm{decline}}=18.5\), and extinction by \(m_{\mathrm{ext}}=52\). For reference, we also track (but do not simulate at) coarser units: a \emph{generation} is taken as three months (90 days). The initial population is \(N_0=8\), and the environmental carrying capacity is \(K=3840\).

\subsection{Uncertainty of rank and its schedule}
We posit that behavioral activation is driven by uncertainty about social rank ('pecking order'). Let \(p_t \in [\tfrac{1}{3},1]\) denote the concentration of mass on a single stratum (dominant / intermediate / subordinate). The mapping from \(p_t\) to an \emph{uncertainty scalar} \(u_t\in[0,1]\) is linear:
\begin{equation}
\label{eq:u-of-p}
u_t \;=\; u(p_t) \;=\; 1 - 1.5\,\Bigl(p_t - \tfrac{1}{3}\Bigr),
\end{equation}
so that \(u(1/3)=1\) (maximal uncertainty), \(u(1/2)=0.75\), and \(u(1)=0\) (no uncertainty).

The time course \(p_t\) captures a three-phase trajectory: constant high uncertainty during growth, gradual partial resolution during the plateau, then a smooth exponential approach to full certainty in decline:
\begin{equation}
\label{eq:p-schedule}
p_t \;=\;
\begin{cases}
\dfrac{1}{3}, & t < m_{\mathrm{plateau}},\\[6pt]
\dfrac{1}{3} + \dfrac{t - m_{\mathrm{plateau}}}{\,m_{\mathrm{decline}} - m_{\mathrm{plateau}}\,}\Bigl(\dfrac{1}{2}-\dfrac{1}{3}\Bigr), & m_{\mathrm{plateau}} \le t < m_{\mathrm{decline}},\\[10pt]
\dfrac{1}{2} + \dfrac{1}{2}\Bigl(1 - e^{-\kappa\,(t - m_{\mathrm{decline}})}\Bigr), & t \ge m_{\mathrm{decline}},
\end{cases}
\qquad \kappa=0.25.
\end{equation}
Equations~\eqref{eq:u-of-p}--\eqref{eq:p-schedule} generate \(u_t\) for use in the game and demographic dynamics below.

\subsection{Game parameterization by uncertainty}
Each month, rats pairwise choose between \emph{engage} (\(C\), standing in for dominance / reproduction / nurturing activities) or \emph{withdraw} (\(D\)). The row player's payoff matrix is made uncertainty-dependent:
\begin{equation}
\label{eq:payoffs}
\begin{array}{c|cc}
 & C & D\\
\hline
C & R(u_t) & S(u_t)\\
D & T(u_t) & P
\end{array}
\qquad
\begin{aligned}
R(u) &= 3 + 2u,\\
T(u) &= 5 - 0.5u,\\
S(u) &= 2.5u,\\
P &= 1.
\end{aligned}
\end{equation}
Thus, as uncertainty rises, mutual engagement becomes more rewarding \((R\uparrow)\), unilateral exploitation becomes slightly less tempting \((T\downarrow)\), and being the sole engager is less punitive \((S\uparrow)\). In the certainty limit \(u=0\), incentives to engage collapse.

\paragraph{Equilibrium engagement rate.}
Let \(x_t\in[0,1]\) denote the (symmetric) equilibrium fraction of agents who choose \(C\) at time \(t\). We use the standard mixed-strategy condition (row indifferent between \(C\) and \(D\)):
\begin{equation}
\label{eq:xstar}
x_t \;=\;
\begin{cases}
1, & R(u_t)\ge T(u_t)\ \text{and}\ S(u_t)\ge P,\\[3pt]
0, & R(u_t)\le T(u_t)\ \text{and}\ S(u_t)\le P,\\[3pt]
\dfrac{P - S(u_t)}{\,R(u_t) - S(u_t) - T(u_t) + P\,}, & \text{otherwise,}
\end{cases}
\end{equation}
with the degenerate denominator case handled by \(x_t=\tfrac{1}{2}\). We also record the complement \(b_t = 1 - x_t\) as the 'beautiful ones' proportion.

\subsection{Population dynamics with capacity and phase effects}
Let \(N_t\) be population at month \(t\). The one-step update is the sum of survivors and newborns:
\begin{equation}
\label{eq:Nt-update}
N_{t+1} \;=\; s_t\,N_t \;+\; r_t\,x_t^{2}\,N_t.
\end{equation}
The survival rate \(s_t\) and reproduction factor \(r_t\) are phase-specific transforms of base parameters \(\bar{s}\) and \(\bar{r}\) (per month), modulated further by overcrowding:
\begin{equation}
\label{eq:phase-rates}
(s_t, r_t) \;=\;
\begin{cases}
\bigl(\bar{s},\ 2\bar{r}\bigr), & t \le m_{\mathrm{plateau}} \quad \text{(growth)},\\[3pt]
\bigl(0.98\,\bar{s},\ 0.6\,\bar{r}\bigr), & m_{\mathrm{plateau}} < t < m_{\mathrm{decline}} \quad \text{(plateau)},\\[3pt]
\Bigl(\bigl(0.98 - 0.38\,a_t\bigr)\bar{s},\ \bigl(0.8 - 0.8\,a_t\bigr)\bar{r}\Bigr), & t \ge m_{\mathrm{decline}} \quad \text{(decline)},
\end{cases}
\end{equation}
where the decline acceleration
\begin{equation}
\label{eq:accel}
a_t \;=\; \biggl(\min\Bigl\{1,\ \frac{t - m_{\mathrm{decline}}}{\,m_{\mathrm{ext}} - m_{\mathrm{decline}}\,}\Bigr\}\biggr)^{5}
\end{equation}
enforces a slow initial deterioration that becomes rapidly catastrophic near extinction.\footnote{As implemented, \(r_t\) decreases from \(0.8\bar{r}\) toward \(0\) by \(t=m_{\mathrm{ext}}\); \(s_t\) decreases from \(0.98\bar{s}\) toward \(0.60\bar{s}\).}

\paragraph{Overcrowding penalty.}
When \(N_t > 0.9K\), we downscale both reproduction and survival to reflect density stress. Let
\begin{equation}
\label{eq:oc-factor}
\phi_t \;=\; \frac{0.9K}{N_t} \quad (\text{applied only if } N_t>0.9K).
\end{equation}
Then
\begin{equation}
\label{eq:oc-apply}
r_t \leftarrow r_t\,\phi_t^{2}, 
\qquad
s_t \leftarrow s_t\,\Bigl(0.5 + 0.5\,\phi_t\Bigr).
\end{equation}
The birth term in \eqref{eq:Nt-update} is proportional to \(x_t^2\) because new litters arise only from mutual engagement \((C,C)\). In the results of the simulations produced this is not activated at any time but introduced as a potential dynamic.

\subsection{Social stratification from engagement level}
We map the monthly engagement rate \(x_t\) to a tripartite stratification \((D_t,I_t,S_t)\) for dominant, intermediate, and subordinate counts (summing to \(N_t\)). First, choose baseline proportions by regime:
\begin{equation}
\label{eq:baseline-strata}
(\pi^{\mathrm{dom}}, \pi^{\mathrm{int}}, \pi^{\mathrm{sub}}) \;=\;
\begin{cases}
\bigl(0.4 + 0.1x_t,\ 0.35,\ 1 - (0.4 + 0.1x_t) - 0.35\bigr), & x_t \ge 0.8,\\[3pt]
(0.35,\ 0.40,\ 0.25), & 0.3 \le x_t < 0.8,\\[3pt]
(0.20,\ 0.30,\ 0.50), & x_t < 0.3,
\end{cases}
\end{equation}
then apply a phase adjustment reflecting mobility in growth and rigidity in decline:
\begin{equation}
\label{eq:phase-bias}
\pi^{\mathrm{dom}} \leftarrow \pi^{\mathrm{dom}} + \delta_t,\qquad
\pi^{\mathrm{sub}} \leftarrow \pi^{\mathrm{sub}} - \delta_t,\qquad
\delta_t =
\begin{cases}
+0.05, & t \le m_{\mathrm{plateau}},\\
0, & m_{\mathrm{plateau}} < t < m_{\mathrm{decline}},\\
-0.05, & t \ge m_{\mathrm{decline}}.
\end{cases}
\end{equation}
Finally renormalize the proportions to sum to one and set
\begin{equation}
(D_t, I_t, S_t) \;=\; \Bigl(\,\big\lfloor \pi^{\mathrm{dom}} N_t \big\rceil,\ \big\lfloor \pi^{\mathrm{int}} N_t \big\rceil,\ N_t - \big\lfloor \pi^{\mathrm{dom}} N_t \big\rceil - \big\lfloor \pi^{\mathrm{int}} N_t \big\rceil \Bigr).
\end{equation}

\subsection{Parameters used}
Unless otherwise noted, simulation results are generated with base monthly parameters \(\bar{r}=0.3\) and \(\bar{s}=0.95\), initial condition \(N_0=8\), carrying capacity \(K=3840\), and phase landmarks \((m_{\mathrm{plateau}},m_{\mathrm{decline}},m_{\mathrm{ext}})=(10.5,\,18.5,\,52)\).

\subsection{Simulation algorithm}
\begin{algorithm}[H]
\caption{Monthly update of uncertainty, behavior, population, and strata}
\begin{algorithmic}[1]
\State \textbf{Input:} \(N_0, K, \bar{r}, \bar{s}, m_{\mathrm{plateau}}, m_{\mathrm{decline}}, m_{\mathrm{ext}}, \kappa\).
\For{$t=0,1,\dots,T-1$}
  \State Compute $p_t$ by \eqref{eq:p-schedule} and $u_t=u(p_t)$ by \eqref{eq:u-of-p}.
  \State Form payoffs $R(u_t),T(u_t),S(u_t),P$ by \eqref{eq:payoffs}.
  \State Compute $x_t$ by \eqref{eq:xstar}; set $b_t=1-x_t$.
  \State Set $(s_t,r_t)$ by \eqref{eq:phase-rates}; if $N_t>0.9K$, apply \eqref{eq:oc-apply}.
  \State Update population by \eqref{eq:Nt-update}: $N_{t+1}= s_t N_t + r_t x_t^2 N_t$, rounded to the nearest integer and truncated at zero if needed.
  \State Compute $(D_t,I_t,S_t)$ from $x_t$ and $t$ via \eqref{eq:baseline-strata}--\eqref{eq:phase-bias}.
\EndFor
\end{algorithmic}
\end{algorithm}

An Agent Based Modeling approach is showcased in section~\ref{sec:ABM} of the Appendix that produced similar results when hierarchy uncertainty is a parameter which drives survival activities.

\section{Results}
\label{sec:results}

\begin{figure}[!htbp]
  \centering
  \includegraphics[width=\linewidth]{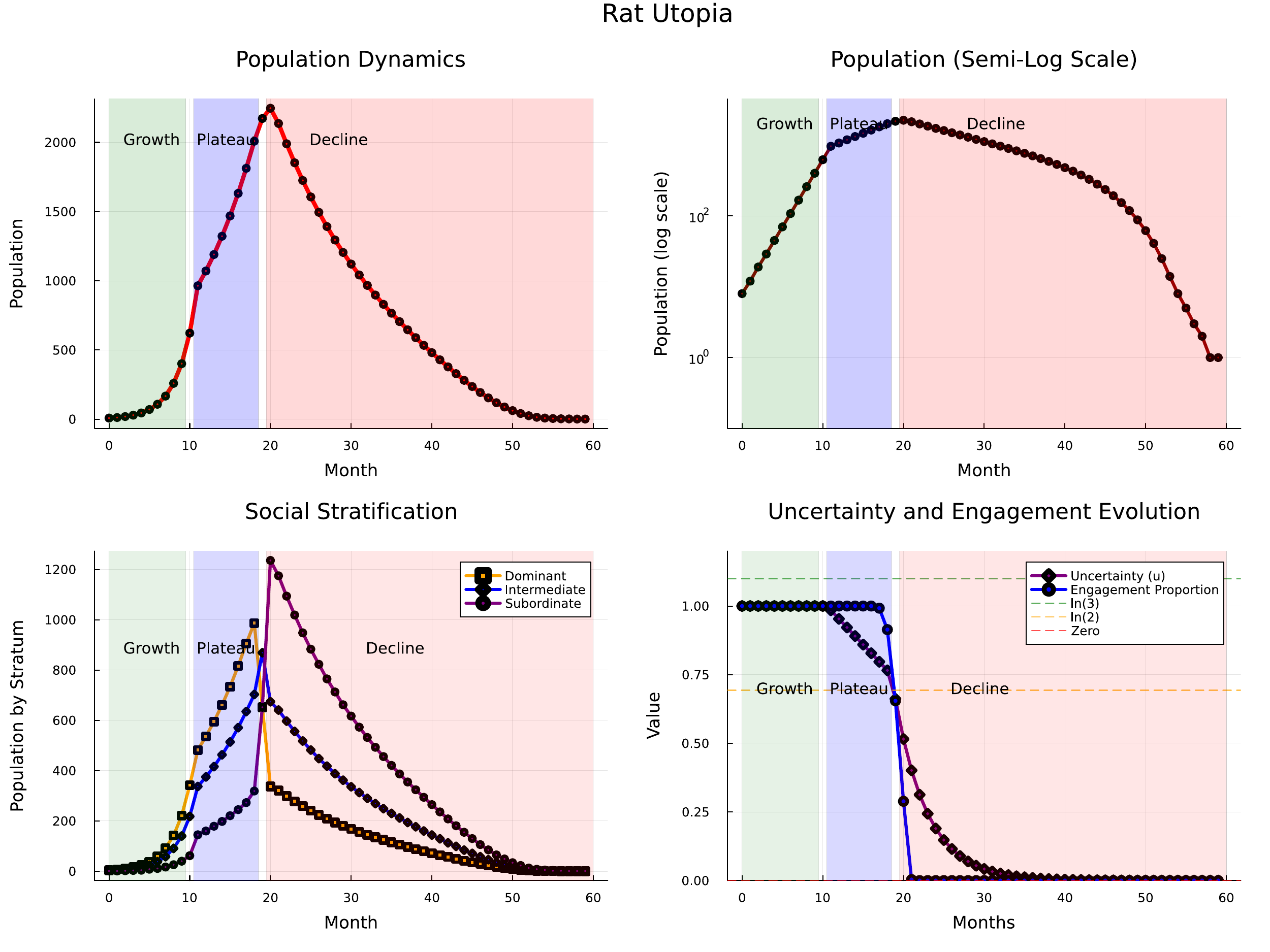}
  \caption{\textbf{Simulation results for the uncertainty-parameterized game model.}
  \emph{(a)} Population on a linear scale with shaded phase bands (growth, plateau, decline) and annotated landmarks; the peak ($\sim 2{,}200$) occurs near month $\sim 20$.
  \emph{(b)} Population on a semi-log scale highlighting an approximately exponential rise in early months, slope decrease across the plateau window, and near-exponential decline after month $18.5$ with late-stage acceleration toward extinction ($\sim 52$ months).
  \emph{(c)} Social stratification (dominant, intermediate, subordinate) inferred from the equilibrium engagement rate $x(t)$ and phase-specific mobility/rigidity; subordinate counts dominate during decline.
  \emph{(d)} Uncertainty $u(t)$ and equilibrium engagement proportion $x(t)$: $u\approx 1$ (and thus $x\approx 1$) during growth, gradual reduction of $u$ across the plateau, and a rapid collapse of $x$ as $u\to 0$ at decline onset, which suppresses births (term proportional to $x^2$ in the population update).}
  \label{fig:matrixGame}
\end{figure}

Figure~\ref{fig:matrixGame} summarizes the simulated population trajectory and internal state of the colony using the game matrix methodology presented. The model reproduces the canonical three-phase pattern reported for Universe~25. In the \emph{growth} phase (months $0$–$10.5$), the population increases approximately exponentially from the initial $N_0=8$ (Panels~a,b), reaching $\sim 6\times10^2$ by the end of this window. This behavior is driven by maximal rank uncertainty ($u\approx 1$) and a correspondingly high equilibrium engagement rate ($x\approx 1$; Panel~d), which together yield a large $(C,C)$ incidence and sustained births. During the \emph{plateau} phase ($10.5$–$18.5$ months), $u(t)$ is programmatically reduced from $1$ toward $0.75$; the induced changes in the payoff matrix lower $x(t)$ modestly, weakening net growth without immediately reversing it. As a result, the total population continues to rise and reaches a peak of $\sim 2{,}200$ near month $\sim 20$ (Panel~a), consistent with the scale of the empirical peak in Universe~25. On the semi-log plot (Panel~b), this transition appears as a reduced growth of the slope, indicating a departure from pure exponential growth.

The \emph{decline} phase (months $\ge 18.5$) begins once uncertainty starts collapsing toward zero. Engagement $x(t)$ falls sharply over a few months (Panel~d), the “beautiful ones” fraction $1-x(t)$ becomes larger, and the birth term—which is proportional to $x^2$—vanishes. With reproduction suppressed and survival simultaneously degraded by the phase-dependent rates and density effects, the population enters a  decay that increases a decay rate on the semi-log axis (Panel~b), with a late-stage acceleration imposed by the worsening survival schedule. Extinction occurs by $\sim 52$ months, again matching the qualitative timing of the historical record. The stratification panel (Panel~c) shows that dominant and intermediate classes expand during growth and early plateau, but subordinate counts dominate the composition throughout decline, reflecting the behavioral withdrawal predicted by the uncertainty-parameterized game.

\section{Discussion}

The results from the game-theoretic model supports the hypothesis that the degeneration of hierarchy entropy driven by the enclosure's full visibility can drive the population collapse observed in the Calhoun Rat Utopia experiments. By parameterizing the payoff matrix with uncertainty levels that decrease over generations, the simulation reproduces the characteristic growth-plateau-decline trajectory, without invoking overcrowding, anxiety/stress, or resource scarcity as primary drivers. High initial uncertainty (maximal entropy) incentivizes engagement in dominance contests, reproduction, and care, yielding rapid population expansion as mutual cooperation dominates the equilibrium. As visibility solidifies the hierarchy, entropy erodes, shifting the equilibrium toward withdrawal and non-engagement, akin to the exploitation dynamics in the visible Prisoner's Dilemma (analyzed in the Appendix). This mechanism aligns with behavioral ecology literature, where uncertainty reduction diminishes motivational signals, such as dopamine responses to novel or ambiguous social cues \cite{bromberg2009midbrain, angela2005uncertainty}, ultimately extinguishing the behavioral fuel for population maintenance.

The model simplifies the hierarchy into three tiers and assumes an uncertainty schedule, it highlights the critical role of environmental design in social dynamics, the enclosure's structure facilitated full symmetric visibility, reinforcing transgenerational hierarchy rigidity. Limitations include the discrete nature of the uncertainty decay.

\appendix

\section{Relationship to Prisoner's Dilemma}
\label{app:PD}

\begin{center}
\begin{tabular}{c|cc}
              & \textbf{C} (\emph{Cooperate}) & \textbf{D} (\emph{Defect}) \\
\hline
\textbf{C} & $(R,\;R)$ & $(S,\;T)$ \\
\textbf{D} & $(T,\;S)$ & $(P,\;P)$
\end{tabular}
\end{center}

\begin{itemize}
  \item $T$ - \textbf{Temptation} (best‐reply payoff when the opponent cooperates).
  \item $R$ - \textbf{Reward} for mutual cooperation.
  \item $P$ - \textbf{Punishment} for mutual defection.
  \item $S$ - \textbf{Sucker's payoff} (worst‐case when you cooperate and the opponent defects).
\end{itemize}

The Prisoner's Dilemma (PD) is a fundamental game in game theory that illustrates the conflict between individual rationality and collective benefit. The payoffs are defined by four key parameters, each representing a specific outcome in the game. These parameters satisfy the inequalities \( T > R > P > S \) (ensuring defection is tempting and dominant individually) and \( 2R > T + S \) (ensuring mutual cooperation is Pareto-optimal over mixed outcomes). The standard numerical values used are:

\begin{itemize}
    \item \( R = 3 \): The \textbf{Reward} for mutual cooperation (both players choose C, or Cooperate). This is the payoff each player receives when both cooperate, representing a beneficial collective outcome. It is higher than P but lower than T to capture the dilemma.
    \item \( S = 0 \): The \textbf{Sucker's payoff} for being exploited (one player cooperates while the other defects, i.e., CD for Player 1 or DC for Player 2). This is the worst outcome for the cooperating player, highlighting the risk of unilateral cooperation.
    \item \( T = 5 \): The \textbf{Temptation} to defect (one player defects while the other cooperates, i.e., CD for Player 2 or DC for Player 1). This is the highest individual payoff, incentivizing defection to exploit the other's cooperation.
    \item \( P = 1 \): The \textbf{Punishment} for mutual defection (both players choose D, or Defect). This is a mediocre outcome, better than S but worse than R, representing the cost of mutual non-cooperation.
\end{itemize}

Players choose actions \( a_i \in \{ C, D \} \) (Cooperate or Defect). The payoffs are symmetric: Player 1's payoff is \( u_1(a_1, a_2) \), and Player 2's is \( u_2(a_1, a_2) = u_1(a_2, a_1) \).
The payoff matrix, with rows as Player 1's choices (C then D) and columns as Player 2's (C then D), is:
\begin{equation}
    \begin{pmatrix}
        (R, R) & (S, T) \\
        (T, S) & (P, P)
    \end{pmatrix}
    =
    \begin{pmatrix}
        (3, 3) & (0, 5) \\
        (5, 0) & (1, 1)
    \end{pmatrix}
\end{equation}
These values are chosen for illustration; the qualitative results hold for any parameters satisfying the inequalities.
Throughout the rest of the paper we study an \emph{iterated} version of this game in which, at the start of each round, \textbf{each} player independently observes the partner’s intended action with probability \(q\in[0,1]\).  All pay-offs in subsequent sections are expectations built from the four events \((1-q)^2,\;q(1-q),\;q(1-q),\;q^{2}\) applied to the matrix above.

\subsection{One-Shot PD Without Visibility}
In the standard one-shot Prisoner’s Dilemma the moves are simultaneous and
unseen, i.e.\ the independent peek probability is \(q=0\).

\subsubsection*{Nash Equilibrium}
A profile \((a_1^*,a_2^*)\) is a Nash equilibrium if no player benefits
from a unilateral deviation:
\[
  u_i(a_i^*,a_{-i}^*) \;\ge\; u_i(a_i,a_{-i}^*) \quad\forall a_i.
\]
Because \(T>R\) and \(P>S\), defection strictly dominates cooperation, so the
unique Nash outcome is \((D,D)\) with pay-offs \((P,P)=(1,1)\).

\subsubsection*{Minimax Value}
The best guarantee a player can secure against the worst response is
\[
  \max_{a_i} \min_{a_{-i}} u_i(a_i,a_{-i}) \;=\;1,
\]
again achieved by choosing \(D\).  Thus the game’s benchmark pay-off is 1,
well below the cooperative level 3.

\subsection{One-Shot PD With Visibility (Reactive Choices)}
\label{subsec:oneshot-visible}

We now allow \emph{transparent intentions}.  
At the start of the round each player independently observes the partner’s intended move with probability \(q\in(0,1]\).  
Thus four observation events are possible:

\begin{center}
\begin{tabular}{lcl}
no one sees                           &:& $(1-q)^2$,\\
only Player 1 sees Player 2           &:& $q(1-q)$,\\
only Player 2 sees Player 1           &:& $q(1-q)$,\\
both see (full mutual transparency)   &:& $q^{2}$.\\
\end{tabular}
\end{center}

A strategy may now condition on whether an intention is observed.

\subsubsection*{Best replies and Nash equilibrium}

\begin{itemize}
\item \emph{Optimal reaction rule.}  
      If a player sees the partner intend \(C\), the best reply is \(D\) (gains \(T=5\)).  
      If a player sees the partner intend \(D\), the best reply is still \(D\) (avoids \(S=0\)).
\item \emph{Implication.}  
      Any intended \(C\) is exploited whenever it is visible, so the only pure-strategy Nash equilibrium is \((D,D)\), yielding pay-offs \((P,P)=(1,1)\) for every \(q>0\).
\end{itemize}

\subsubsection*{Expected pay-offs under attempted cooperation}

Suppose, for illustration, that both players \emph{intend} to cooperate
but follow the optimal reaction rule once they observe a move.

\begin{center}
\begin{tabular}{lccc}
\toprule
Event & Probability & Final outcome & Pay-off to Player 1 \\ \midrule
no one sees           & $(1-q)^2$       & $CC$ & $R=3$ \\
only P1 sees P2’s \(C\)& $q(1-q)$        & $DC$ & $T=5$ \\
only P2 sees P1’s \(C\)& $q(1-q)$        & $CD$ & $S=0$ \\
both see               & $q^{2}$         & $DD$ & $P=1$ \\ \bottomrule
\end{tabular}
\end{center}

\[
\mathbb{E}[u_1]
  =(1-q)^2\cdot 3
   + q(1-q)\cdot 5
   + q(1-q)\cdot 0
   + q^{2}\cdot 1
  = 3 - q - q^{2}.
\]

Hence
\(\mathbb{E}[u_1]=3\) at \(q=0\) and falls monotonically to
\(\mathbb{E}[u_1]=1\) at full transparency \(q=1\).  
The same holds for Player 2 by symmetry.

\paragraph{Take-away.}
Visibility enables instantaneous exploitation, so the cooperative
pay-off ceiling drops from \(3\) to \(3-q-q^{2}\) and reaches the
punishment level \(1\) when the game is fully transparent
(\(q=1\)).  Correspondingly, the unique Nash equilibrium for any
positive \(q\) is \((D,D)\).

\subsection{One Shot Prisoner's Dilemma comparison with and without Visibility}

\begin{table}[!htbp]
\centering
\caption{Key metrics in a single‐round PD for two information regimes.
         Visibility is symmetric and independent: each player sees the
         partner’s intended move with probability \(q\in[0,1]\).}
\label{tab:pd-visible-oneshot}
\renewcommand{\arraystretch}{1.15}
\begin{tabular}{@{}p{4.3cm}p{4.1cm}p{4.6cm}@{}}
\toprule
\textbf{Metric} &
Hidden actions (no visibility, q=0) &
Transparent intentions (visibility, q>0) \\ \midrule
Nash-equilibrium strategy & \((D,D)\) & \((D,D)\) \\[2pt]
Equilibrium pay-off (each) & \(1\) & \(1\) \\[2pt]
Expected pay-off \emph{if both intend to co-operate} & \(3\) & \(3-q-q^{2}\) \\[2pt]
Max–min (secure) value & \(1\) & \(1\) \\[2pt]
Effect on co-operation & No exploitation; but \(D\) still dominates. &
Seeing intentions allows immediate exploitation; the cooperative
pay-off ceiling falls from \(3\) to \(3-q-q^{2}\), reaching
\(1\) at full transparency (\(q=1\)). \\ \bottomrule
\end{tabular}
\end{table}

\noindent
\textit{Interpretation.} Introducing even partial visibility does not
create new equilibria; it simply lowers the best achievable symmetric
pay-off when players \emph{try} to co-operate, while the Nash outcome
remains mutual defection in either case.

\subsection{Iterated PD Without Visibility (simultaneous moves)}
We first recall the \emph{classical} repeated Prisoner’s Dilemma in which
moves are simultaneous and never observed in advance
(\(q=0\)).  Each round-\(t\) payoff \(u_i^{t}\) is
discounted by \(\delta\in(0,1)\); a player’s long-run value is
\((1-\delta)\sum_{t\ge0}\delta^{t}u_i^{t}\).
Because every action is publicly revealed \emph{after} the round, strategies
can depend on the full history of past outcomes.

\subsubsection*{Equilibria and the folk-theorem}
\begin{itemize}
\item \textbf{Stage-game equilibrium.}  The unique Nash profile is
      \((D,D)\) with pay-off \(1\).
\item \textbf{Folk theorem.}  If players are sufficiently patient
      (\(\delta\) high), \emph{any} feasible pay-off above the minmax
      level 1 can be sustained in a subgame-perfect equilibrium.  
      In particular, perpetual co-operation \((3,3)\) is enforceable
      whenever
      \[
        \delta \;\ge\; \frac{T-R}{T-P}
                   \;=\; \frac{5-3}{5-1}
                   \;=\; 0.5.
      \]
\item \textbf{Canonical strategies.}
      \begin{description}
      \item[Grim trigger]  Co-operate until the first defection, then
            defect forever; works iff \(\delta\ge0.5\).
      \item[Tit-for-Tat]   Start with \(C\); thereafter copy the
            partner’s previous move.  More forgiving, but still requires
            high \(\delta\) for stability.
      \end{description}
\end{itemize}

\subsubsection*{Incentive calculation (grim trigger)}
Let \(V_C = \dfrac{3}{1-\delta}\) be the value of perpetual
co-operation.  A one-shot deviation yields
\(V_D = 5 + \dfrac{\delta}{1-\delta}\).
The no-deviation condition \(V_C \ge V_D\) reduces to
\(\delta \ge 0.5\).

When moves remain hidden during the round, patient players can maintain
the efficient path \((3,3)\) through the threat of future punishment.
The benchmark pay-off therefore rises from the one-shot level 1 to the
full 3 as soon as \(\delta\) exceeds 0.5.  This “high-trust” outcome
serves as the yard-stick for the visibility cases analysed next.

\subsection{Iterated PD \emph{with} symmetric, independent visibility}
Each round proceeds as in the one–shot transparent game: after choosing
their \emph{intended} action, both players independently observe the
partner’s intention with probability \(q\in(0,1]\).  
Final moves and pay-offs are then revealed to both sides.

\subsubsection*{Effective stage–game pay-offs}
If the intention profile is \((C,C)\) the four observation events give

\[
\begin{array}{lcl}
\text{nobody sees}      &:& (1-q)^2      \;\Rightarrow\; (3,3),\\
\text{only 1 sees}      &:& q(1-q)       \;\Rightarrow\; (5,0),\\
\text{only 2 sees}      &:& q(1-q)       \;\Rightarrow\; (0,5),\\
\text{both see}         &:& q^{2}        \;\Rightarrow\; (1,1).
\end{array}
\]

Hence the expected pay-off from mutual \(C\) is \(3-q-q^{2}\).
The unique one-shot Nash equilibrium remains \((D,D)\) with pay-off 1.

\subsubsection*{Co-operation under imperfect private monitoring}

The game now fits the “imperfect monitoring” version of the folk theorem.

\begin{itemize}
\item \textbf{Noisy signal.}  Deviating to \(D\) is not always detected
      — it produces the same \((5,0)\) or \((0,5)\) signal that a
      bad-luck peek would create under intended \(C\).
\item \textbf{Simple grim trigger fails.}  Punishing after the first
      non-\((3,3)\) round leads to many false punishments when \(q>0\).
\item \textbf{Review strategies.}  A player can instead punish only if
      the “both defect” signal \((1,1)\) appears, or if too many
      exploitations accumulate in a review block.  Standard results
      (e.g.\ Kandori \& Matsushima 1998) show that \emph{near-efficient}
      pay-offs up to \(3-q-q^{2}\) are sustainable provided
      \(\delta\) is close to 1 and \(q\) is not too large.
\end{itemize}

For every fixed \(q<1\) there exists \(\bar\delta(q)\uparrow1\) such that
all pay-offs strictly above 1 and below \(3-q-q^{2}\) belong to a 
subgame-perfect equilibrium whenever \(\delta\ge\bar\delta(q)\).
At full transparency \(q=1\) the ceiling collapses to 1 and only
\((D,D)\) survives.

Partial visibility makes cheating easier \emph{and} detection more
credible.  Trust can still emerge, but only if (i) peeking is rare
enough and (ii) the group tolerates occasional mistakes by requiring
several bad signals before retaliation.  As visibility becomes common
or the future is discounted heavily, pay-offs slide back to the
self-protective baseline of mutual defection.

\subsection{Pay-off summary}

\begin{itemize}
\item \textbf{One-shot game}.  
      Whether moves are hidden \((q=0)\) or fully transparent \((q=1)\), rational players defect, so the pay-off is fixed at \(1 \;\longrightarrow\; 1\).

\item \textbf{Iterated game without visibility \((q=0)\).}  
      When the same pair meet repeatedly and value the future highly
      (\(\delta\gtrsim0.5\)), the threat of later punishment keeps
      co-operation stable; the long-run pay-off rises to the full
      \(3\).

\item \textbf{Iterated game with partial visibility \((0<q<1)\).}  
      Peeking erodes the co-operative surplus: even for patient players
      the best symmetric pay-off is capped at
      \(3-q-q^{2}\;\bigl(\le 3\bigr)\).

\item \textbf{Full visibility \((q=1)\).}  
      Every intended move is seen in advance; exploitation is certain,
      so the iterated pay-off collapses to the baseline \(1\).

\item \textbf{Rule of thumb.}  
      More rounds \emph{and} less peeking push the average reward
      upward toward \(3\); frequent peeking wipes the gain and puts the
      game back at \(1\).
\end{itemize}

\subsection{Simulation results}

To visualize how the average pay-off depends jointly on (i) the length of
the relationship and (ii) the probability of seeing a partner’s intended
move, we ran a Monte-Carlo experiment:

\begin{itemize}
\item \textbf{Game.}  Iterated Prisoner’s Dilemma with independent,
      symmetric visibility~\(q\in[0,1]\).  
\item \textbf{Strategy pair.}  \emph{Tit-for-Tat with opportunistic
      defection}: start with \(C\); in later rounds intend to copy the
      opponent’s \emph{final} action; if a \(C\) intention is observed
      switch to \(D\) for that round.
\item \textbf{Grid.}  Peek probability
      \(q=\{0,0.05,\dots,1\}\) and horizon
      \(T=\{1,5,10,\dots,100\}\).
\item \textbf{Metric.}  Mean per-round pay-off for one player,
      averaged over \(300\) independent games for each \((q,T)\) pair.
\end{itemize}

\begin{figure}[!htbp]
  \centering
  \includegraphics[width=.75\linewidth]{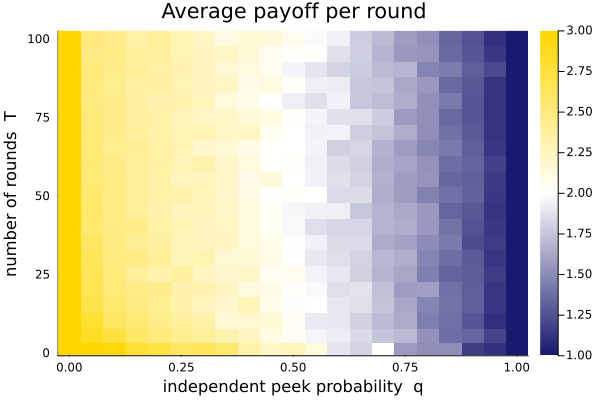}
  \caption{\label{fig:payoff1}%
    Mean pay-off per round as a function of visibility probability \(q\)
    (horizontal axis) and number of rounds \(T\) (vertical axis).  
    Colour scale is fixed at the theoretical bounds:
    dark~blue = punishment level \(1\); bright yellow = full
    co-operation \(3\). This uses Tit-For-Tat and opportunistic defection.}
\end{figure}

When moves are almost never seen (\(q\approx0\)) the pay-off rises from
\(1\) in the one-shot case \(T=1\) toward the co-operative maximum \(3\)
as the horizon lengthens.  
Even a modest peek probability (\(q\approx0.3\)) trims that surplus: the
bright band fades, capping the reward near \(3-q-q^{2}\simeq2.4\).  
Beyond \(q\approx0.8\) the surface turns dark irrespective of \(T\); the
game effectively collapses to mutual defection and the pay-off locks at
the baseline \(1\).  
The heat-map therefore matches the theoretical summary: \emph{more
rounds and less peeking expand the space for trust, whereas frequent
peeking erases the gain and pushes the population back to the
self-protective equilibrium}.
Hidden moves fail because no one can trust the other to cooperate; visible moves fail because trust is instantly exploitable.
Now  it can be asked whether a similar ‘visibility kills trust’ effect appears in animal dominance interactions—specifically, how rats decide whether to risk conflict or courtship when they can observe each other’s intentions.”

\section{Equation based Modeling Approach}
\label{app:EquationModel}
The premise of hierarchy and uncertainty in the rat utopia model is integrated into the differential equations as follows: The hierarchy represents the social pecking order, which evolves from chaotic (high uncertainty) in early generations to rigidly fixed (low uncertainty) in later ones. Uncertainty is quantified via entropy \(H\), which decreases as the rats gain a 'better idea' of the hierarchy reflecting knowledge from conflicts and distance from the center due to visibility and certainty in their positions. This aligns with the description:

\begin{itemize}
    \item Generation 0: Rats thrown in, maximum uncertainty (uniform probability \(1/3\) across three potential strata: e.g., top, middle, bottom), so \(H = -\sum_{i=1}^{3} \frac{1}{3} \ln\left(\frac{1}{3}\right) = \ln(3) \approx 1.099\) (in natural units, or 'nats').
    \item Generation 1: Still forming, but better idea; uncertainty remains at \(1/3\) across three strata, so \(H \approx 1.099\).
    \item Generation 2: Hierarchy clarifies (e.g., uncertainty now uniform over two effective strata: dominant vs. subordinate), so \(H = -\sum_{i=1}^{2} \frac{1}{2} \ln\left(\frac{1}{2}\right) = \ln(2) \approx 0.693\).
    \item Generation 3: Perfect visibility (probability \(1/1\) for each rat's fixed position), so \(H = 0\).
\end{itemize}
This entropy \(H\) parameterizes the effort equation, where high \(H\) (early uncertainty) drives effort \(E\) upward, motivating rats to fight/reproduce to alter the hierarchy. As \(H\) decreases (hierarchy becomes visible and fixed), effort drops, especially for rats with high social position \(S\) (top of hierarchy, less need to strive). The population equation couples this to growth: high early \(E\) causes a boom, moderate \(E\) a plateau, and low \(E\) a crash---mirroring how visibility stifles productive behavior, akin to defection in a visible prisoner's dilemma.

The equations are:
\begin{align}
\label{eq:Model1a}
\frac{dE}{dt} &= k \, H \, (1 - S) - d(H) \, E, \\
\label{eq:Model1b}
\frac{dP}{dt} &= r \, E \, P \left(1 - \frac{P}{K}\right) - \left(m + \alpha \left(1 - \frac{H}{\ln 3}\right)\right) P,
\end{align}
where:
- \(d(H) = d_{\text{base}}\) if \(H > 0\), else \(d(H) = d_{\text{collapse}}\) (piece-wise to enforce sudden demotivation at full visibility),
- \(E\) is the average effort level for dominance/reproduction,
- \(P\) is the population size,
- \(H\) is the entropy of pecking order uncertainty (decreasing over generations: \(\ln(3) \to \ln(2) \to 0\)),
- \(S\) is the average social position (0 for bottom, 1 for top),
- Parameters: \(k=1.0\), \(r=0.2\), \(K=2400\), \(d_{\text{base}}=0.05\), \(d_{\text{collapse}}=1.0\), \(m=0.015\), \(\alpha=0.15\).

Generational shifts in \(H\) are approximated via time (months): high \(H\) for early chaos (\(t < 15\), generations 0--1), medium for strata formation (\(15 \leq t < 30\), generation 2), zero for full visibility (\(t \geq 30\), generation 3) timed to align with overlapping litters and density-driven visibility in the experiment.

\begin{figure}[!htbp]
  \centering
  \includegraphics[width=1.0\linewidth]{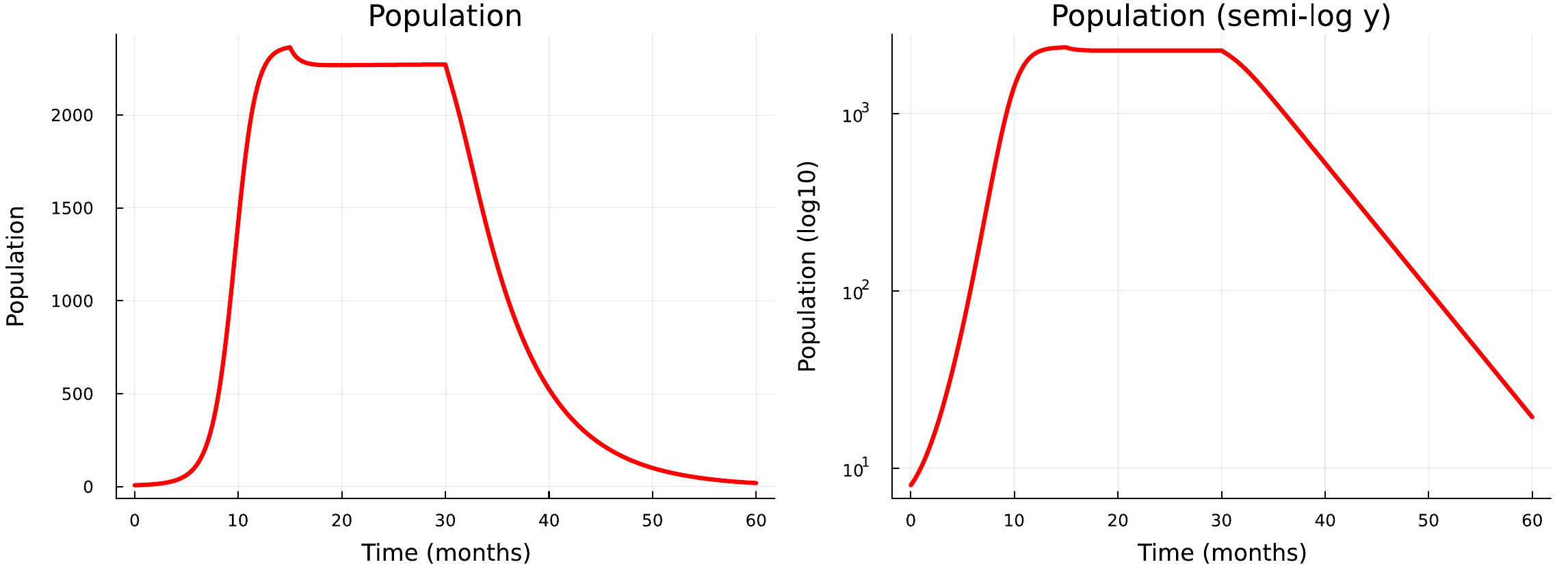}
  \caption{\label{fig:Model1}%
    Results of running a simulation using the modeling paradigm defined in Eq~\ref{eq:Model1a} and Eq~\ref{eq:Model1b}.}
\end{figure}
Fig~\ref{fig:Model1} shows the results from the simulation of the differential equation model defined in Eq~\ref{eq:Model1a} and Eq~\ref{eq:Model1b}. The distinct phases and general population pattern match over time can be seen. Although the slight declining slope in the plateau phase is not displayed and the collapse phase should have a gradually decreasing trend the overall structure is evident. The equations can be tuned to match the experiments as needed and the parameter values used for a deeper interpretation.

\section{Rat Utopia Agent Based Model approach}
\label{sec:ABM}

To simulate Calhoun’s Universe as a monthly, discrete-time agent-based system, each rat $i$ has attributes
$\{\text{sex},\,\text{age},\,\text{rank}\in\{\text{dominant},\text{intermediate},\text{subordinate}\},\,\text{propensity},\,\text{alive}\}$ and inhabits a single well-mixed compartment of capacity $K$.
Macro-level uncertainty is driven by a piecewise schedule $p(t)$ (growth: $p=1/3$; plateau: linear $1/3\!\to\!1/2$; decline: $1/2\!\to\!1$) that is mapped to
\[
u(t)=1-1.5\bigl(p(t)-\tfrac{1}{3}\bigr)\in[0,1].
\]
Individual engagement (the drive to participate in social/reproductive activity) is
$e_i(t)=\sigma\!\big(k\big(u(t)-u_0\big)\big)\cdot m_{\text{rank}}(\text{rank}_i)$
with $\sigma(x)=1/(1+e^{-x})$, where $m_{\text{rank}}$ mildly boosts dominants and attenuates subordinates.
Survival uses phase-specific baselines $s_{\text{phase}}(t)$, a crowding term $s_{\text{crowd}}(N_t,K)$ (penalized only when $N_t>0.9K$), and small age/rank modifiers (infant risk, senescence, subordinate stress in decline).
Ranks evolve via an uncertainty- and phase-dependent Markov rule with upward mobility in growth, weak mobility in plateau, and downward drift in decline.
A late-decline behavioral subtype (“beautiful ones”) is flagged when $u(t)$ is low and a subordinate’s $e_i(t)$ remains persistently low.

Mating is female-driven (polygynous): each engaged adult female conceives independently with probability
$q(t)=q_{\text{phase}}(t)\,r_{\text{crowd}}(N_t,K)\,r_{\text{density}}(N_t)$ provided at least one engaged adult male exists. Litter size
$L\sim\mathrm{Poisson}(\lambda_{\text{phase}}(t))$ (forced $L\ge1$ outside the decline phase). Newborns receive sex at $0.5$ probability and a birth rank drawn from a distribution that shifts with $u(t)$, reducing the bias toward subordination during high-uncertainty growth. After updating survival, engagement, reproduction, and rank mobility, we record population, mean engagement, and rank counts, then advance the month; the run terminates at extinction or a fixed horizon.

\begin{algorithm}[!htbp]
\caption{Uncertainty-driven ABM for Calhoun’s Universe}
\label{alg:rat-abm}
\begin{algorithmic}[1]
\State Initialize $N_0$ agents (balanced sex), ages, and ranks; set capacity $K$ and horizon $T$.
\For{$t=0$ to $T-1$}
  \State Compute $p(t)$, $u(t)=1-1.5(p(t)-\tfrac{1}{3})$, and phase $\in\{\text{Growth, Plateau, Decline}\}$.
  \ForAll{agents $i$}
    \State $\text{age}_i \leftarrow \text{age}_i+1$; if not alive, continue.
    \State $s \leftarrow s_{\text{phase}}(t)\cdot s_{\text{crowd}}(N_t,K)\cdot s_{\text{mod}}(i,\text{phase})$; \textbf{if} $\mathrm{Uniform}(0,1)>s$ \textbf{then} kill $i$.
    \State $e_i \leftarrow \sigma\!\big(k(u(t)-u_0)\big)\cdot m_{\text{rank}}(\text{rank}_i)$; set $\text{engaging}_i \sim \mathrm{Bernoulli}(e_i)$.
    \State $\text{rank}_i \leftarrow \mathrm{Mobility}\big(\text{rank}_i, u(t), \text{phase}\big)$.
    \State \textbf{if} (phase$=$Decline $\land$ rank$_i=$Subordinate $\land\ e_i<\tau$) \textbf{then} mark ``beautiful''.
  \EndFor
  \State $M \leftarrow$ engaged adult males; $F \leftarrow$ engaged adult females.
  \If{$M\neq\emptyset$}
    \State $q \leftarrow q_{\text{phase}}(t)\cdot r_{\text{crowd}}(N_t,K)\cdot r_{\text{density}}(N_t)$; $\lambda \leftarrow \lambda_{\text{phase}}(t)$.
    \ForAll{$f\in F$}
      \If{$\mathrm{Bernoulli}(q)=1$}
        \State Draw $L\sim \mathrm{Poisson}(\lambda)$; enforce $L\ge 1$ if phase $\neq$ Decline.
        \For{$\ell=1$ to $L$}
          \State sex $\sim \mathrm{Bernoulli}(0.5)$; rank $\sim \mathrm{BirthRank}(u(t))$.
          \State Add newborn $(\text{sex},\,\text{age}=0,\,\text{rank})$.
        \EndFor
      \EndIf
    \EndFor
  \EndIf
  \State Record $(N_{t+1},\,u(t),\,\bar{e}(t),\,\text{rank counts})$; advance to $t{+}1$ or stop if $N_{t+1}=0$.
\EndFor
\end{algorithmic}
\end{algorithm}

\begin{figure}[!htbp]
  \centering
  \includegraphics[width=\textwidth]{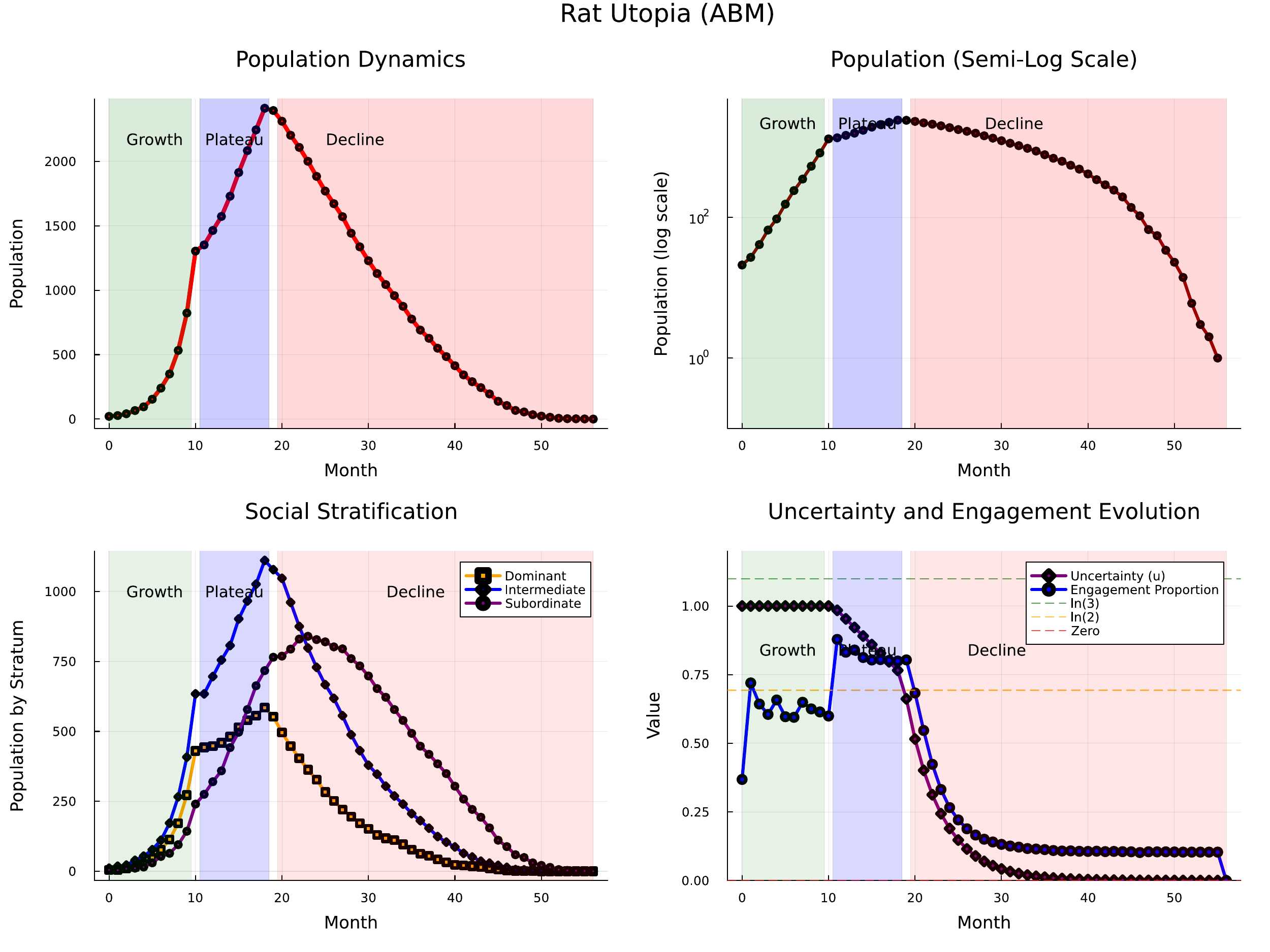}
  \caption{Agent-based simulation of Calhoun’s universe driven by a time-varying uncertainty schedule. 
  Panels show (top-left) population on a linear scale, (top-right) population on a semi-log scale, 
  (bottom-left) social stratification counts (dominant, intermediate, subordinate), and 
  (bottom-right) macro-uncertainty $u(t)$ with the mean engagement proportion $\bar e(t)$. 
  Shaded bands indicate the Growth, Plateau, and Decline phases. 
  With polygynous, female-driven conception and phase-dependent survival, the population rises to 
  $\sim\!10^3$ by the end of Growth, peaks near $\sim\!2\times10^3$ late in Plateau, and then collapses 
  to extinction by month $\approx 52$ as $u(t)\!\to\!0$ and engagement vanishes.}
  \label{fig:rat-abm}
\end{figure}

Figure~\ref{fig:rat-abm} summarizes the uncertainty-driven ABM. The semi-log panel clarifies the 
approximate exponential rise during Growth, a shallower slope through Plateau (due to reduced conception 
probability and a soft density throttle), and a rapid exponential-like decay in Decline. The stratification 
panel shows upward mobility and a larger intermediate/dominant contingent during high-uncertainty periods, 
followed by a reversal toward subordination as uncertainty falls and mobility tightens. The engagement curve 
tracks $u(t)$ by design, providing the mechanistic link between the macro-level uncertainty schedule and the 
observed demographic boom–bust dynamics.

\begin{thebibliography}{23}
\providecommand{\natexlab}[1]{#1}
\providecommand{\url}[1]{\texttt{#1}}
\expandafter\ifx\csname urlstyle\endcsname\relax
  \providecommand{\doi}[1]{doi: #1}\else
  \providecommand{\doi}{doi: \begingroup \urlstyle{rm}\Url}\fi

\bibitem[Angela and Dayan(2005)]{angela2005uncertainty}
J~Yu Angela and Peter Dayan.
\newblock Uncertainty, neuromodulation, and attention.
\newblock \emph{Neuron}, 46\penalty0 (4):\penalty0 681--692, 2005.

\bibitem[Appleby(1983)]{appleby1983probability}
Michael~C Appleby.
\newblock The probability of linearity in hierarchies.
\newblock \emph{Animal Behaviour}, 31\penalty0 (2):\penalty0 600--608, 1983.

\bibitem[Beery and Kaufer(2015)]{beery2015stress}
Annaliese~K Beery and Daniela Kaufer.
\newblock Stress, social behavior, and resilience: insights from rodents.
\newblock \emph{Neurobiology of stress}, 1:\penalty0 116--127, 2015.

\bibitem[Bromberg-Martin and Hikosaka(2009)]{bromberg2009midbrain}
Ethan~S Bromberg-Martin and Okihide Hikosaka.
\newblock Midbrain dopamine neurons signal preference for advance information
  about upcoming rewards.
\newblock \emph{Neuron}, 63\penalty0 (1):\penalty0 119--126, 2009.

\bibitem[Broom and Cannings(2002)]{broom2002modelling}
M~Broom and C~Cannings.
\newblock Modelling dominance hierarchy formation as a multi-player game.
\newblock \emph{Journal of theoretical Biology}, 219\penalty0 (3):\penalty0
  397--413, 2002.

\bibitem[Calhoun(1962)]{calhoun1962population}
John~B Calhoun.
\newblock Population density and social pathology.
\newblock \emph{Scientific American}, 206\penalty0 (2):\penalty0 139--149,
  1962.

\bibitem[Calhoun(1973)]{calhoun1973death}
John~B Calhoun.
\newblock Death squared: The explosive growth and demise of a mouse population,
  1973.

\bibitem[Calhoun(1977)]{calhoun1977looking}
John~B Calhoun.
\newblock \emph{Looking Backward from" the Beautiful Ones"}.
\newblock US Department of Health, Education and Welfare, National Institutes
  of Health, 1977.

\bibitem[Chen et~al.(2023)Chen, Winiarski, Pu{\'s}cian, Knapska, Walczak, and
  Mora]{chen2023generalized}
Xiaowen Chen, Maciej Winiarski, Alicja Pu{\'s}cian, Ewelina Knapska,
  Aleksandra~M Walczak, and Thierry Mora.
\newblock Generalized glauber dynamics for inference in biology.
\newblock \emph{Physical Review X}, 13\penalty0 (4):\penalty0 041053, 2023.

\bibitem[Chen et~al.(2025)Chen, Winiarksi, Pu{\'s}cian, Knapska, Mora, and
  Walczak]{chen2025modeling}
Xiaowen Chen, Maciej Winiarksi, Alicja Pu{\'s}cian, Ewelina Knapska, Thierry
  Mora, and Aleksandra~M Walczak.
\newblock Modeling collective behavior in groups of mice housed under
  semi-naturalistic conditions.
\newblock \emph{eLife}, 13:\penalty0 RP94999, 2025.

\bibitem[Clutton-Brock et~al.(1986)Clutton-Brock, Albon, and
  Guinness]{clutton1986great}
TH~Clutton-Brock, SD~Albon, and FE~Guinness.
\newblock Great expectations: dominance, breeding success and offspring sex
  ratios in red deer.
\newblock \emph{Animal behaviour}, 34\penalty0 (2):\penalty0 460--471, 1986.

\bibitem[Dall et~al.(2005)Dall, Giraldeau, Olsson, McNamara, and
  Stephens]{dall2005information}
Sasha~RX Dall, Luc-Alain Giraldeau, Ola Olsson, John~M McNamara, and David~W
  Stephens.
\newblock Information and its use by animals in evolutionary ecology.
\newblock \emph{Trends in ecology \& evolution}, 20\penalty0 (4):\penalty0
  187--193, 2005.

\bibitem[Donelson et~al.(2018)Donelson, Salinas, Munday, and
  Shama]{donelson2018transgenerational}
Jennifer~M Donelson, Santiago Salinas, Philip~L Munday, and Lisa~NS Shama.
\newblock Transgenerational plasticity and climate change experiments: Where do
  we go from here?
\newblock \emph{Global Change Biology}, 24\penalty0 (1):\penalty0 13--34, 2018.

\bibitem[Eccard et~al.(2017)Eccard, Dammhahn, and Yl{\"o}nen]{eccard2017bruce}
Jana~A Eccard, Melanie Dammhahn, and Hannu Yl{\"o}nen.
\newblock The bruce effect revisited: is pregnancy termination in female
  rodents an adaptation to ensure breeding success after male turnover in low
  densities?
\newblock \emph{Oecologia}, 185\penalty0 (1):\penalty0 81--94, 2017.

\bibitem[Elwood(2009)]{elwood2009difficulties}
Robert Elwood.
\newblock Difficulties remain in distinguishing between mutual and
  self-assessment in animal contests.
\newblock \emph{Animal Behaviour}, 2009.

\bibitem[Hock and Huber(2006)]{hock2006modeling}
Karlo Hock and Robert Huber.
\newblock Modeling the acquisition of social rank in crayfish: winner and loser
  effects and self-structuring properties.
\newblock \emph{Behaviour}, pages 325--346, 2006.

\bibitem[Kraeuter et~al.(2018)Kraeuter, Guest, and Sarnyai]{kraeuter2018open}
Ann-Katrin Kraeuter, Paul~C Guest, and Zolt{\'a}n Sarnyai.
\newblock The open field test for measuring locomotor activity and anxiety-like
  behavior.
\newblock In \emph{Pre-clinical models: techniques and protocols}, pages
  99--103. Springer, 2018.

\bibitem[Marsden(1972)]{marsden1972crowding}
Halsey~M Marsden.
\newblock Crowding and animal behavior.
\newblock 1972.

\bibitem[Perony et~al.(2012)Perony, K{\"o}nig, and
  Schweitzer]{perony2012stochastic}
Nicolas Perony, Barbara K{\"o}nig, and Frank Schweitzer.
\newblock A stochastic model of social interaction in wild house mice.
\newblock \emph{arXiv preprint arXiv:1212.0662}, 2012.

\bibitem[Ramsden(2009)]{ramsden2009urban}
Edmund Ramsden.
\newblock The urban animal: population density and social pathology in rodents
  and humans, 2009.

\bibitem[Ramsden and Adams(2009)]{ramsden2009escaping}
Edmund Ramsden and Jon Adams.
\newblock Escaping the laboratory: the rodent experiments of john b. calhoun \&
  their cultural influence.
\newblock \emph{Journal of Social History}, pages 761--792, 2009.

\bibitem[Shizuka and McDonald(2015)]{shizuka2015network}
Daizaburo Shizuka and David~B McDonald.
\newblock The network motif architecture of dominance hierarchies.
\newblock \emph{Journal of the Royal Society Interface}, 12\penalty0
  (105):\penalty0 20150080, 2015.

\bibitem[van Haeringen and Hemelrijk(2022)]{van2022hierarchical}
Erik van Haeringen and Charlotte Hemelrijk.
\newblock Hierarchical development of dominance through the winner-loser effect
  and socio-spatial structure.
\newblock \emph{PLoS One}, 17\penalty0 (2):\penalty0 e0243877, 2022.

\end{thebibliography}

\end{document}